\documentclass[twocolumn,english,aps,prd,nofootinbib]{revtex4-2}
\usepackage{lmodern}
\usepackage[T1]{fontenc}
\usepackage[latin9]{inputenc}
\usepackage{xcolor}
\usepackage{babel}
\usepackage{amsmath}
\usepackage{amssymb}
\usepackage{graphicx}
\usepackage[unicode=true]
 {hyperref}

\makeatletter

\usepackage{enumerate}

\makeatother

\begin{document}
\title{Time-reversed information flow through a wormhole in scalar--tensor
gravity}
\author{Hoang Ky Nguyen$\,$}
\email[\ \ ]{hoang.nguyen@ubbcluj.ro}

\affiliation{Department of Physics, Babe\c{s}--Bolyai University, Cluj-Napoca
400084, Romania}
\author{Francisco S. N. Lobo$\,$}
\email[\ \ ]{fslobo@fc.ul.pt}

\affiliation{Instituto de Astrof\'{\i}sica e Ci\^{e}ncias do Espa\c{c}o, Faculdade
de Ci\^encias da Universidade de Lisboa, Campo Grande, Edif\'{\i}cio
C8, P-1749-016 Lisbon, Portugal}
\affiliation{Departamento de F\'{\i}sica, Faculdade de Ci\^encias da Universidade
de Lisboa, Campo Grande, Edif\'{\i}cio C8, P-1749-016 Lisbon, Portugal}
\date{July 2, 2024}
\begin{abstract}
{\vskip2pt}This Letter aims to advance unexplored properties of a
new class of Closed Timelike Curves recently discovered in scalar--tensor
gravity, reported in \textcolor{purple}{\href{https://doi.org/10.3390/universe9110467}{Universe {\bf 9}, 467 (2023)}}
and \textcolor{purple}{\href{https://doi.org/10.1140/epjc/s10052-023-11805-3}{Eur.$\,$Phys.$\,$J.$\,$C {\bf 83}, 626 (2023)}}.
Therein, it was shown that when the Weak Energy Condition is violated, the topology of spacetime in scalar--tensor gravity is altered, enabling the formation of two-way
traversable wormholes. Furthermore, each of these wormholes acts a
gateway between two \emph{time-mirrored} worlds,
where the two asymptotically flat sheets in the Kruskal--Szekeres
diagram are glued antipodally along \emph{three} directions---time
$t$ and the polar and azimuth angles $(\theta,\,\varphi)$ of the
2-sphere---to form a wormhole throat. This contrasts with the standard
embedding diagram which typically glues the sheets only along the
$\theta$ and $\varphi$ directions. Crucially, due to the `gluing'
along the $t$ direction, the wormhole becomes a portal connecting
the two spacetime sheets with \emph{opposite} physical time flows,
enabling the emergence of closed timelike loops which straddle the
throat. In this Letter, we shall point out that this portal \emph{mathematically}
permits the possibility of backward propagation of information \emph{against}
time. This feature is ubiquitous for wormholes in scalar--tensor
theories. In addition, we formulate the Feynman sum for transition
amplitudes of microscopic particles in the proximity of a wormhole
throat in which we account for timelike paths that experience time
reversal.
\end{abstract}
\maketitle
\emph{Introduction}---Closed timelike curves (CTCs) are mathematical
constructs first emerging in the context of General Relativity via
the pioneering works of van Stockum and G\"odel \citep{Stockum-1938,Godel}.
These curves represent paths through spacetime that loop back on themselves,
allowing an object or observer to return to an event in its own past.
Interests in CTCs were revived in the works of Thorne and Novikov
in the context of the Morris--Thorne wormholes which are thought
to occur in the presence of exotic matter (or in modified gravity)
\citep{MorrisThorne-1988-1,MorrisThorne-1988-2,NovikovCTC,frolovnovikovTM,Echeverria,Visser-book}.
Despite the problematic nature of CTCs, the mathematical validity
sparked considerable interest among gravitational theorists \citep{Tipler-CTCs,Pfarr,Ori:2005ht,Ori,Alcubierre,Alcubierre:2017pqm,CauchyCTC,EverettCTC,Felice,GottCTC,hawking}.
The concept of CTCs serves as a ``gedanken experiment'', compelling
us to grapple with and reevaluate the foundational principles of general
relativity and its modifications.\vskip4pt

\begin{figure}[!t]
\noindent \begin{centering}
\includegraphics[scale=0.75]{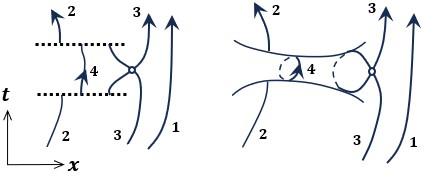}\vspace{-.25cm}
\par\end{centering}
\caption{\label{fig:DP-surgery}The Deutsch--Politzer surgery. In the left
panel, a 2-dimensional Minkowski spacetime is incised along the two
dotted segments. The banks of the cuts are then glued together to
form a `handle', as shown in the right panel. Paths 1 and 2 represent
regular trajectories in Minkowski spacetime. Path 3 is self-crossing,
resulting in a portion of it forming a closed timelike curve (CTC).
Path 4 depicts an infinite loop.}
\end{figure}

Soon after the works of Thorne and collaborators, Deutsch and Politzer
attempted to investigate quantum effects associated with CTCs. They
introduced the Deutsch--Politzer (DP) space---an abstract construct
created through surgical manipulation on a two-dimensional Lorentzian
spacetime, forming a cylinder-shaped `handle' that bridges two spacelike
regions \citep{Deutsch-1991,Politzer-1992,Krasnikov-1995,Krasnikov-1998}.
This process is visually depicted in Fig.$\ $\ref{fig:DP-surgery},
where Paths 3 and 4 represent CTCs. As an object moves along a CTC,
proper time continues to cumulate (i.e., the time count registered
by a clock carried along the path keeps increasing), but the object
eventually returns to its original spacetime point. Whereas geometry
pertains to local properties of spacetime, topology pertains to global
ones; in crafting a `handle', the DP surgery deliberately alters the
\emph{topology} of spacetime, allowing for CTCs to emerge.\vskip4pt

While DP space, as a conceptual model, serves its purpose, it is preferable
to \emph{derive} alterations to spacetime topology from a concrete
theory of gravity. In fact, in General Relativity, it was shown that
through the Einstein field equations these spacetimes violate the
Weak Energy Condition (WEC) \citep{MorrisThorne-1988-1}. Matter that
violates the WEC is denoted \textit{exotic matter}, and can spontaneously
induce spacetime to acquire non-trivial topologies, such as the Morris-Thorne
wormholes that facilitates CTCs \citep{MorrisThorne-1988-2,Visser-book,Harko:2013yb,frolovnovikovTM,Bronnikov-2010}.\vskip4pt

In a series of recent papers \citep{Nguyen-2022-Lambda0,Nguyen-2022-Buchdahl,Nguyen-2023-essay},
one of the authors extended Buchdahl's work in pure $\mathcal{R}^{2}$
gravity \citep{Buchdahl-1962}, culminating in a static spherically symmetric vacuum solution. Notably,
despite the absence of exotic matter, the higher-derivative terms
in the theory formally violate the WEC, under certain conditions.
This violation enables the construction of a two-way traversable wormhole,
which we named the Morris--Thorne--Buchdahl (MTB) wormhole \citep{2023-WH}.
Importantly, the analytical solution has allowed us to construct a
($\zeta-$)Kruskal--Szekeres (KS) diagram, revealing a non-trivial
topology. The $(\zeta-$)KS coordinates enable for an identification
procedure that results in the wormhole throat bridging two asymptotically
flat spacetime sheets with opposing flows of proper
time, as we recently proposed in \citep{2023-CTC}. Physical processes
occurring in the two spacetime sheets across the throat evolve in
opposite time directions. In essence, the MTB wormhole throat serves
as a ``gateway to another \emph{time-reversed} world''. More generally,
we have found that this property is a universal feature for Brans--Dicke
and scalar--tensor gravities \citep{2023-WEC}. It is important to
note that embedding diagrams, a standard tool to visualize wormholes,
are\linebreak \emph{inadequate} in exposing this feature, as we shall
explain in a later section of this Letter.\vskip8pt

Importantly, the construction of the CTC that we presented in \citep{2023-CTC}
\emph{does not necessitate one wormhole mouth to move at high speed
or be situated near a supermassive object to accumulate time dilation}---a
procedure commonly portrayed in the literature \citep{MorrisThorne-1988-2,Visser-book}.
The resulting CTC is of a novel type, offering intriguing avenues
for exploration and analysis.\vskip12pt

\begin{figure}[!t]
\noindent \begin{centering}
\includegraphics[scale=0.45]{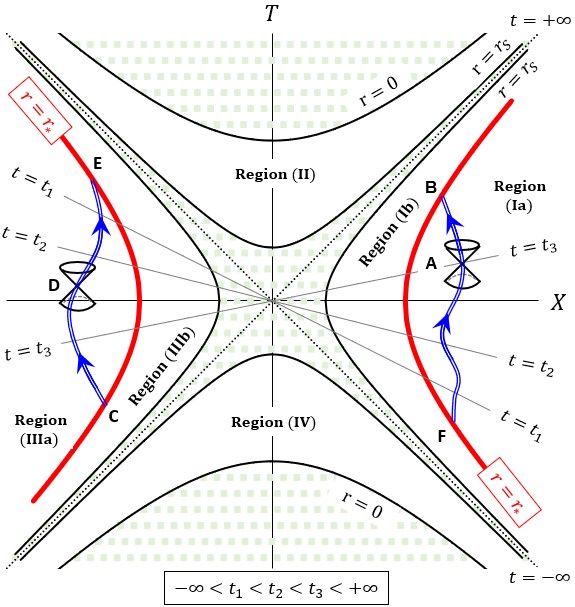}\vspace{0.2cm}
\par\end{centering}
\noindent \begin{centering}
\hspace{-.5cm}\includegraphics[scale=0.5]{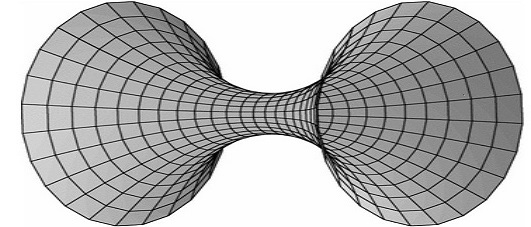}\vspace{0cm}
\par\end{centering}
\caption{\label{fig:KS-diagram} Visualizing a Buchdahl-inspired spacetime.
While the embedding diagram (lower panel) suggests `gluing' at a fixed
timeslice $T$, the $\zeta-$KS diagram (upper panel) makes the `gluing'
along the \emph{time} direction, with the two thick red lines are
identical \emph{antipodally}. An MTB wormhole is formed by flipping
Region (IIIa) upside down and gluing it with Region (Ia).\linebreak
Trajectory (ABCDEFA) forms a closed timelike loop.}
\end{figure}

\emph{The special Buchdahl-inspired metric}---Consider the vacuum
pure $\mathcal{R}^{2}$ field equation \citep{Nguyen-2022-Lambda0}
\begin{equation}
\mathcal{R}\,\Bigl(\mathcal{R}_{\mu\nu}-\frac{1}{4}g_{\mu\nu}\mathcal{R}\Bigr)+g_{\mu\nu}\square\,\mathcal{R}-\nabla_{\mu}\nabla_{\nu}\mathcal{R}=0\label{eq:R2-field-eqn}
\end{equation}
Among a variety of solutions \citep{Nguyen-2022-Buchdahl,2023-WEC},
the \emph{asymptotically flat} Buchdahl-inspired solution to Eq. \eqref{eq:R2-field-eqn}
expressed in the KS coordinates is given by \citep{Nguyen-2022-Lambda0}\vskip-6pt

\small
\begin{equation}
ds^{2}=-4e^{-r^{*}(r)}\Bigl(1-\frac{r_{\text{s}}}{r}\Bigr)^{\frac{\tilde{k}+1}{\zeta}}\left(dT^{2}-dX^{2}\right)+\Bigl(1-\frac{r_{\text{s}}}{r}\Bigr)^{\frac{\zeta+\tilde{k}-1}{\zeta}}r^{2}d\Omega^{2}\label{eq:KS-metric}
\end{equation}
{\normalsize}and\vspace{-.5cm}
\begin{align}
T^{2}-X^{2} & =-e^{r^{*}(r)};\ \ \ \frac{T}{X}=\tanh\frac{t}{2r_{\text{s}}}\label{eq:T-X}
\end{align}
which depend on a dimensionless (Buchdahl) parameter $\tilde{k}$,
and $\zeta:=\left(1+3\,\tilde{k}^{2}\right)^{1/2}$, whereas the tortoise
coordinate $r^{*}(r)$ involves a Gaussian hypergeometric function
and is given in \citep{Nguyen-2022-Lambda0}. Note: per Eqs. \eqref{eq:T-X},
each pair of coordinates $(t,r)$ corresponds to \emph{two} pairs
of KS coordinates $(T,X)$ and $(-T,-X)$.\vskip4pt

In Ref.$\,$\citep{2023-WH} we showed that, for $\tilde{k}\in(-1,0)$,
the WEC is formally violated. Simultaneously, the areal radius, $R(r)=r\,\Bigl(1-\frac{r_{\text{s}}}{r}\Bigr)^{\frac{\zeta+\tilde{k}-1}{2\,\zeta}}$,
attains a (local) minimum at 
\begin{equation}
r_{*}=\frac{\zeta-\tilde{k}+1}{2\,\zeta}\,r_{\text{s}}\label{eq:r-star}
\end{equation}
which exceeds $r_{\text{s}}$ when $\tilde{k}\in(-1,0)$. In this
scenario, the spacetime forms a two-way traversable wormhole, referred
to as the MTB wormhole, with its throat located at $r_{*}$ given
in \eqref{eq:r-star}, to be briefly described below. An embedding
diagram is shown in Fig.$\ $\ref{fig:KS-diagram}.\vskip4pt

Restricting within the $(T,X)$ plane (viz. $d\theta=d\varphi=0$),
the $\zeta-$KS diagram for the metric \eqref{eq:KS-metric}--\eqref{eq:T-X}
is shown in Fig.$\ $\ref{fig:KS-diagram}. The diagram is self-explanatory
\citep{2023-WH}: (i) It is conformally Minkowski, with null geodesics
being $dX=\pm dT$. (ii) A constant--$t$ contour corresponds to
a straight line running through the origin of the $(T,X)$ plane.
(iii) Regions (I) and (III) are mirror images of each other, upon
flipping the sign of the KS coordinates, viz. $(T,X)\leftrightarrow(-T,-X)$.
(iv) For $\tilde{k}\in(-1,0)$, the hyperbola $r=r_{*}$ (thick red
lines) splits these regions into sub-regions (Ia), (Ib), (IIIa), (IIIb).
The `outer' sub-regions (Ia) and (IIIa) are asymptotically flat spacetime
sheets which are devoid of physical singularities.\vskip12pt
\begin{figure}[!t]
\noindent \begin{centering}
\includegraphics[scale=0.8]{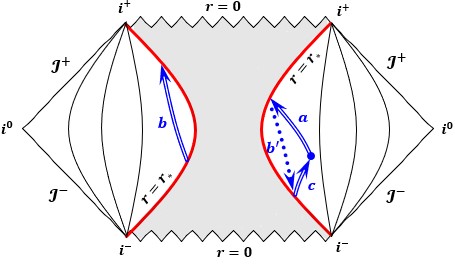}\vspace{-.25cm}
\par\end{centering}
\caption{\label{fig:Penrose-diagram}Penrose diagram illustrating an MTB wormhole
with its two sheets presented as the left and right wedges. Constant-$r$
contours connecting $i^{-}$ with $i^{+}$ are shown. The wedges are
glued along the thick red lines, corresponding to the throat at $r=r_{*}$,
but with the left wedge and line flipped upside down, resulting in
a non-trivial topology where the physical time direction flips across
the throat. Path $(abc)$ is a CTC, with $(b')$ mirroring $(b)$.
The shaded area is not a part of the wormhole; it contains physical
singularities at $r=0$ and $r=r_{\text{s}}$.}
\end{figure}

\emph{Identification of antipodal points and emergence of CTCs}---Antipodal
points play a crucial role in the $\zeta-$KS diagram. Specifically,
the two thick red lines on the diagram, in Fig.$\ $\ref{fig:KS-diagram},
represent a wormhole throat, and the identification of antipodal points
$(T,X)$ and $(-T,-X)$ on these lines signifies \emph{a single spacetime
event occurring on the throat}. It is important to note that this
identification is limited strictly to points on the throat. Points
that stay off the throat do \emph{not} enjoy this privilege; they
correspond to independent events occurring on separate spacetime sheets.
(Note: Our procedure contrasts with historical maneuvers known as
``elliptic/antipodal identification'' \citep{Domenech-1988,Sanchez-1987a,Sanchez-1987b,Rindler-1965,Poplawski-2010,Schrodinger-1957,tHooft-2016,Strauss-2020}.
In particular, when dealing with the regular KS diagram of a Schwarzschild
spacetime, Rindler \citep{Rindler-1965} assigned each antipodal pair
to the same event unrestrictedly, whereas Pop\l{}awski and other
authors \citep{Domenech-1988,Sanchez-1987a,Sanchez-1987b,Poplawski-2010,tHooft-2016,Strauss-2020}
applied such an assignment to antipodal points that lie on the Schwarzschild
horizon.)\vskip4pt

In Fig.$\ $\ref{fig:KS-diagram}, the trajectory $A\rightarrow B\equiv C\rightarrow D\rightarrow E\equiv F\rightarrow A$
constitutes a closed timelike loop. Note that Points B and C, as well
as Points E and F, are antipodal on the throat and, therefore, identical.
The segments between these points can freely take any form as long
as they reside within the lightcones. One key feature is that while
the segment $F\rightarrow A\rightarrow B$ progresses in an increasing
order of the `time' coordinate $t$, the other segment $C\rightarrow D\rightarrow E$
advances in a \emph{decreasing} order of $t$. (Note: the proper time
$\tau$ \emph{increases} along $C\rightarrow D\rightarrow E$. Essentially, the wormhole throat serves as a gateway to another \emph{time-reversed}
world by connecting two sheets with \emph{opposing} physical time
flows. \vskip4pt

Akin to the DP space which incurs an alteration of spacetime topology,
the $\zeta-$KS diagram acquires a non-trivial topology. However,
rather than relying on abstract manipulation as in the DP space, the
change in topology for our $\zeta-$KS diagram takes place spontaneously,
driven by the inherent higher-derivative nature of the $\mathcal{R}^{2}$
theory. The pertinent sections of the Penrose diagram are shown in
Fig.$\ $\ref{fig:Penrose-diagram}, with explanations provided in
its caption. Owing to the antipodality, the left wedge must be flipped
upside down to facilitate the gluing of the two thick red lines in
forming an MTB wormhole.\vskip4pt

It is worth noting that the opposite time flows for physical processes
in Quadrants (I) and (III) have been known since the conception of
the KS diagram in the 1960s. Yet, this fact has often been overlooked
in wormhole studies, most likely because of the prevalent use of embedding
diagrams, which are a standard technique for visualizing wormholes.
Embedding diagrams obscure a fundamental feature, as they suggest
`gluing' the two sheets along the two angular directions, namely the
polar and azimuth angles ($\theta$ and $\varphi$), at the minimal
areal radius. However, in an MTB wormhole, the `gluing' actually occurs
in \emph{three} directions -- the angular ones ($\theta$ and $\varphi$)
\emph{and} the time direction ($t$). This only becomes evident with
the explicit use of the $\zeta-$KS diagram. In Fig. \ref{fig:KS-diagram},
one of the thick red lines needs be flipped upside down, and then
the two said lines get glued in their \emph{entirety}, viz. along
the time direction as well, besides $\theta$ and $\varphi$.\vskip8pt

The portal connecting two time-reversed sheets is ubiquitous not only
in pure $\mathcal{R}^{2}$ gravity but also in Brans--Dicke and scalar--tensor
theories, as they share a similar KS diagram \citep{2023-WEC}. It
is intuitive to expect that time-reversed portals are a common feature
among wormholes in general. It is important to emphasize that the
violation of the WEC plays a crucial role in forming wormholes in
scalar--tensor gravity. In this situation, whereas the (local) geometry
of spacetime remains Riemannian, \emph{its (global) topology has been
drastically altered}, which is a prerequisite for the emergence of
wormholes and CTCs. This fulfills the vision of Deutsch--Politzer
and Krasnikov mentioned in the Introduction \citep{Deutsch-1991,Politzer-1992,Krasnikov-1995,Krasnikov-1998}.\vskip8pt

We will now present three corollaries arising from the existence of
CTCs in the MTB wormhole.\vskip8pt

\begin{figure}[!t]
\noindent \begin{centering}
\includegraphics[scale=0.35]{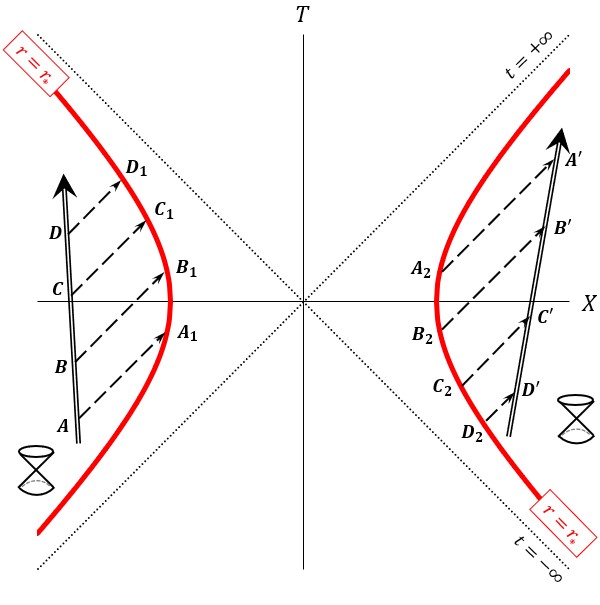}
\par\end{centering}
\noindent \begin{centering}
$\ $\includegraphics[scale=0.35]{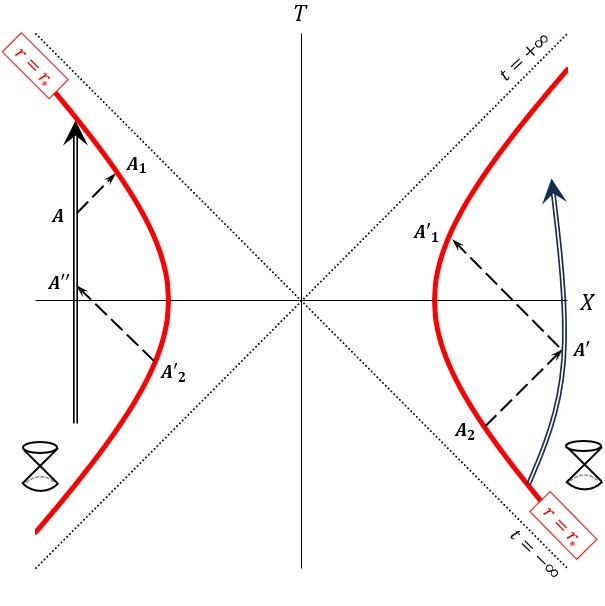}\vspace{-.3cm}
\par\end{centering}
\caption{\label{fig:Corollaries}Illustration of Corollaries \#1 and \#2. Upper
panel: Traveler in the left sheet regularly sends light signals to
Observer in the right sheet (Note: $A_{1}\equiv A_{2}$, $B_{1}\equiv B_{2}$
and so on.) The order of arrival for light signals is \emph{reversed}.
Lower panel: Traveler at $A$ sends out a light signal, then crosses
through the throat. Later on, Traveler while passing by $A'$ deflects
the light signal back; the light signal arrives at $A''$ which precedes
its emission event $A$. (Note: $A_{1}\equiv A_{2}$ and $A'_{1}\equiv A'_{2}$.)}
\end{figure}

\emph{Corollary \#1: Perception of time reversal via light signal
exchange}---The upper panel of Fig.$\ $\ref{fig:Corollaries} exhibits
the case. Consider a Traveler in the left sheet emitting light signals
into the wormhole in the order $A$, $B$, $C$, and $D$; the light
signals emerge into the right mouth and continue on their respective
null paths. Note that on the throat, each antipodal pair correspond
to a single event, viz. $A_{1}\equiv A_{2}$, $B_{1}\equiv B_{2}$,
$C_{1}\equiv C_{2}$, and $D_{1}\equiv D_{2}$. An Observer residing
in the right sheet receives light signals in a \emph{reversed} order,
$D'$, $C'$, $B'$, then $A'$, however. Therefore, from the Observer's
perspective, the Traveler appears to move backward in time. The perception
is mutually symmetric: the Traveler would perceive the Observer as
moving backward with respect to the Traveler.\vskip8pt

\emph{Corollary \#2: Backward flow of information against time}---In
the lower panel of Fig. \ref{fig:Corollaries}, we depict another
closely related effect. Consider a Traveler in the left sheet. At
point $A$, he emits a light signal that enters the wormhole at $A_{1}$
and emerges at $A_{2}$ (note their antipodal nature). The Traveler
then passes into the wormhole and escapes to the right sheet. Choosing
a trajectory such that at point $A'$ (where he crosses paths with
the original light signal), he deflects the light signal back into
the wormhole (e.g., using a mirror). The chosen path bends `outward'
to the right, delaying the reflection event. The deflected light signal
at $A'$ re-enters the throat at $A'_{1}$ and re-emerges at $A'_{2}$
(again, note their antipodal nature). Finally, the light signal arrives
at point $A''$, preceding the emission event $A$.\vskip8pt

\emph{Corollary \#3: Bidirectional time flow on the throat}---A Traveler
sitting precisely on the throat could proceed either forward into
the right sheet or backward into the left sheet. If choosing to reside
on the throat, both directions of time are admissible for him.\pagebreak

\emph{Prevention of free will to act on foreknowledge?}---There seems
to be no inherent principle forbidding microscopic objects from ``exploiting
foreknowledge to their advantage''. Inanimate objects lack free will
to alter history, such as murdering their younger selves or preemptively
securing \emph{the} winning lottery tickets. A timelike path involving
`time travel' for \emph{microscopic} particles would pose no such
threat and should be allowed to contribute in the Feynman sum of transition
amplitudes for these particles, as we will formulate shortly. The
implications of foreknowledge on macroscopic animate objects constitute
a subject of ongoing debate and deliberation, nevertheless \citep{Tipler-CTCs,Pfarr,Ori:2005ht,Ori,NovikovCTC,Alcubierre,Alcubierre:2017pqm,CauchyCTC,Echeverria,EverettCTC,Felice,Godel,GottCTC,hawking}.
We venture one view: macroscopic objects might undergo a strong decoherence
process upon passing through the wormhole throat. If violent enough,
the decoherence could erase, for instance, the mental state of an
intrepid traveler crossing the throat. For example, in Corollary \#2
above, memory erasure could derail the Traveler from carrying out
his plan to deflect the light signal at $A'$, thus thwarting its
`early' arrival at $A''$. This decoherence-based reasoning could
be applicable to macroscopic objects in general,\linebreak such as
an unmanned spaceship pre-programmed to traverse to $A'$ to deflect
the light signal. The decoherence might erase its computer program
and sabotage its planned itinerary.\vskip4pt

Interestingly, in parallel with our decoherence-based proposal that
forbids the use of foreknowledge as discussed herein, a more recent
article has suggested a mechanism to ``erase'' the memory of animate
travelers, potentially avoiding time-travel paradoxes \citep{Gavassino-2024}.\vskip12pt

\emph{Quantum effects induced by time-reversed paths---}Using metric
\eqref{eq:KS-metric}, we can cast the proper time between two events
$A$ and $B$, $\tau_{AB}=\int_{A}^{B}d\tau$, as a functional of
$\vec{r}(t):=\{r(t),\,\theta(t),\,\varphi(t)\}$, given by\small
\begin{align}
\tau_{AB} & =\int_{t_{A}}^{t_{B}}dt\,\Bigl(1-\frac{1}{r}\Bigr)^{\frac{\tilde{k}-1}{2\,\zeta}}\times\label{eq:tau-AB}\\
 & \ \ \ \Bigl[\Bigl(1-\frac{1}{r}\Bigr)^{\frac{2}{\zeta}}-\Bigl(\dot{r}^{2}+\Bigl(1-\frac{1}{r}\Bigr)r^{2}\bigl(\dot{\theta}^{2}+\sin^{2}\theta\,\dot{\varphi}^{2}\bigr)\Bigr)\Bigr]^{\frac{1}{2}}\nonumber 
\end{align}
{\normalsize}with the overdot indicating a derivative with respect
to $t$. Denoting $z:=1-r_{\text{s}}/r\in[r_{*},1)$ with $z_{*}:=1-r_{\text{s}}/r_{*}$
and $r_{*}$ given in Eq.$\ $\eqref{eq:r-star} specifying the location
of the throat, Eq.$\ $\eqref{eq:tau-AB} yields\small
\begin{equation}
\tau_{AB}=\int\displaylimits_{t_{A}}^{t_{B}}\frac{dt\text{\,}z^{\frac{\tilde{k}-1}{2\,\zeta}}}{(1-z)^{2}}\Bigl[z^{\frac{2}{\zeta}}(1-z)^{4}-\dot{z}^{2}-z(1-z)^{2}\bigl(\dot{\theta}^{2}+\sin^{2}\theta\,\dot{\varphi}^{2}\bigr)\Bigr]^{\frac{1}{2}}\label{eq:tau}
\end{equation}
\normalsize

The geodesics in the plane $\theta=\pi/2$, obtained by extremizing
the functional \eqref{eq:tau}, satisfies the following equations,
with $l$ and $\mathcal{E}$ being two constants of motion:
\begin{align}
\dot{z} & =(1-z)^{2}\,z^{\frac{1}{\zeta}}\frac{1}{\mathcal{E}}\sqrt{\mathcal{E}^{2}-(1-z)^{2}z^{\frac{2-\zeta}{\zeta}}l^{2}-z^{\frac{\tilde{k}+1}{\zeta}}}\label{eq:geodesic-z}\\
\dot{\varphi} & =(1-z)^{2}\,z^{\frac{2}{\zeta}-1}\frac{l}{\mathcal{E}}\label{eq:geodesic-varphi}
\end{align}
In the large $\mathcal{E}$ limit, Eqs.$\ $\eqref{eq:geodesic-z}--\eqref{eq:geodesic-varphi}
are integrable to yield $t-t_{0}=\frac{1}{1-1/\zeta}\,z^{1-1/\zeta}\,_{2}F_{1}\left(2,1-1/\zeta;2-1/\zeta;z\right)$
and $\varphi=\text{const}$, corresponding to a pure radial motion.
\vskip4pt
\begin{figure*}[!t]
\noindent \begin{centering}
\includegraphics[scale=0.6]{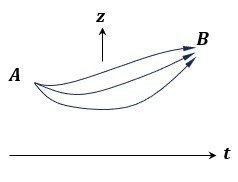}\hspace{1cm}\includegraphics[scale=0.6]{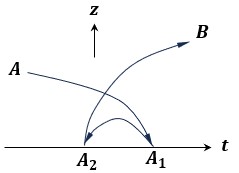}\vspace{.25cm}
\par\end{centering}
\noindent \begin{centering}
\hspace{-.4cm}\includegraphics[scale=0.6]{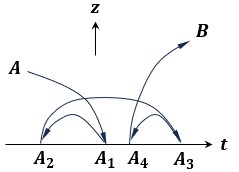}\includegraphics[scale=0.6]{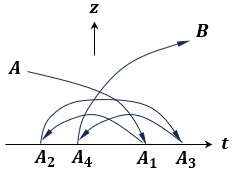}\includegraphics[scale=0.6]{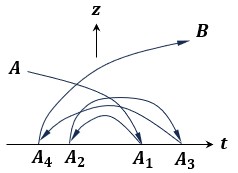}\includegraphics[scale=0.6]{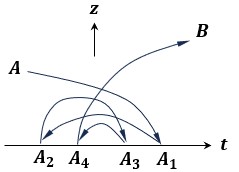}\includegraphics[scale=0.6]{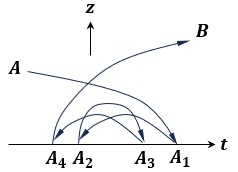}
\par\end{centering}
\caption{\label{fig:path-integral}Different transition paths from Event A
to Event B (both located in the right sheet of the wormhole). In the
upper row, 3 paths exhibit no crossing into the left sheet, while
another path involves a single crossing into the left sheet and subsequent
return. The lower row presents 5 admissible arrangements for paths
with two crossings into the left sheet and twice-returning. In all
cases, traversal in the left sheet is depicted by a segment oriented
against the direction of $t$, meaning that a time-reversed path entails
a \emph{multi-valued} function $z(t)$.}
\end{figure*}

Consider now a pair of starting and end points $A$ and $B$ situated
in the right sheet. Note that Eq.$\ $\eqref{eq:tau} has been defined
for a \emph{single-valued} function $z(t)$ that connects $A$ with
$B$ while staying above the line $z=z_{*}$. This scenario is depicted
in the upper left plot in Fig.$\ $\ref{fig:path-integral}, showing
3 paths with no crossing into the left sheet (the boundary of which
is the horizontal axis $t$, representing $z=z_{*}$).\vskip4pt

When a path crosses the wormhole throat and ventures into the left
sheet, the respective segment in the left sheet has a reversed time
direction, however. This can conveniently be superimposed in the same
$(t,z)$ plane except with an orientation against the direction of
$t$. Other plots in Fig.$\ $\ref{fig:path-integral} showcase the
once-time-reversed and twice-time-reversed paths. Given the additive
nature of proper time, it is straightforward to see, for example,
that for the once-time-reversed path: $\tau_{AA_{1}A_{2}B}=\tau_{AA_{1}}+\tau_{A_{1}A_{2}}+\tau_{A_{2}B}$.\vskip4pt

The transition amplitude of a particle of mass $m$, as usual, is
given by a Feynman sum, $K(\vec{z}_{B},\vec{z}_{A};t_{B}-t_{A})\simeq\int\mathcal{D}\vec{z}(t)\,\exp\left(-i\,m\,\tau_{AB}(\vec{z}(t))\right)$.
Note that in the absence of a wormhole throat, all paths that contribute
to the Feynman sum must run forward in time; this means that the function
$z(t)$ is single-valued. However, when facilitated by a wormhole,
a time-reversed path would make $z(t)$ a \emph{multi-valued} function,
as is evident from Fig.$\ $\ref{fig:path-integral}. The contribution
of time-reversed paths into the Feynman sum is thus a novel non-trivial
aspect. Moreover, in the proximity of a throat, timelike paths with
multiple time reversals could proliferate and significantly---if
not fundamentally---alter the transition amplitude of free particles.
This intuitive possibility underscores the importance of the decoherence
process discussed in relation to safeguarding causality for macroscopic
objects. We leave this arena open for future exploration.\vskip8pt

Consider now the evolution of a system, for simplicity, restricted
within the plane $\theta=\pi/2$. The action of an object of mass
$m$ is $S=-m\int d\tau$. By virtue of the definition of the Lagrangian
via $S=\int dt\,L$, and from Eq.$\ $\eqref{eq:tau}, we deduce the
following Lagrangian 
\begin{equation}
L:=-\frac{m\,z^{\frac{\tilde{k}-1}{2\,\zeta}}}{(1-z)^{2}}\left[z^{\frac{2}{\zeta}}(1-z)^{4}-\dot{z}^{2}-z(1-z)^{2}\,\dot{\varphi}^{2}\right]^{\frac{1}{2}}
\end{equation}
and, via the generalized momenta $p_{z}:=\frac{\delta L}{\delta\dot{z}}$
and $p_{\varphi}:=\frac{\delta L}{\delta\dot{\varphi}}$, the associated
Hamiltonian is given by 
\begin{align}
H & =p_{z}\,\dot{z}+p_{\varphi}\,\dot{\varphi}-L\\
 & =z^{\frac{1}{\zeta}}\sqrt{(1-z)^{4}\,p_{z}^{2}+z^{-1}(1-z)^{2}\,p_{\varphi}^{2}+m^{2}\,z^{\frac{\tilde{k}-1}{\zeta}}}\label{eq:Hamiltonian}
\end{align}
with $p_{\varphi}$ and $H$ being two constants of motion.\vskip4pt

Hamiltonians are known to be suitable for the Heisenberg picture when
describing the evolution of free particles and for the interaction
picture when building the Feynman rules for quantum field theories.
Despite the apparent asymmetric treatment of the time direction (versus
the spatial ones) by the Hamiltonian, relativistic covariance is duly
restored in the final results \citep{Sakurai-AQM}. In the context
of an MTB wormhole, where the background metric is static and inert,
the Hamiltonian could find utility. As a by-product, the form of the
Hamiltonian in Eq.$\ $\eqref{eq:Hamiltonian} is also evocative of
Dirac's historical approach towards the (relativistic) Dirac equation
in 1927. Starting from the energy-momentum relation $E^{2}=p^{2}+m^{2}$
and requiring the field equation for the electron to be first-order,
Dirac conceived the spinors and the associated Clifford algebra for
the Dirac gamma matrices in Minkowski spacetime. Notwithstanding the
tetrad formalism, Expression$\ $\eqref{eq:Hamiltonian} suggests
pathways for an alternative and convenient representation of Dirac's
spinor fields in the presence of an MTB wormhole. The Clifford algebra
may be adapted to conform with the altered energy-momentum relation,
obtained by `squaring' Eq.$\ $\eqref{eq:Hamiltonian}:
\begin{equation}
z^{-\frac{2}{\zeta}}\,E^{2}=(1-z)^{4}\,p_{z}^{2}+z^{-1}(1-z)^{2}\,p_{\varphi}^{2}+m^{2}\,z^{\frac{\tilde{k}-1}{\zeta}}
\end{equation}

Lastly, the time-reversed information flows across a wormhole, as
demonstrated in Corollary \#2, could potentially impact the unitarity
of the evolution operator and the time-ordered product in the interaction
picture. These intriguing prospects warrant in-depth exploration.\vskip8pt

As our final remark, although CTCs appear to be an inevitable feature
of wormholes constructed from scalar--tensor gravity, the decoherence
effect discussed herein, together with the ``memory eraser'' mechanism
independently proposed in \citep{Gavassino-2024}, could prevent paradoxes
related to time travel of animate objects, thereby rescuing scalar--tensor
gravity from inherent inconsistencies.\vskip12pt

\emph{Acknowledgments}---We wish to thank the anonymous referee for
their constructive and valuable comments which helped improve the
quality of our Letter.\linebreak HKN thanks Tiberiu Harko, Mustapha Azreg-A\"inou,
Luis A. Correa-Borbonet, Lorenzo Gavassino, and Nicholas Buchdahl.
FSNL acknowledges support from the Funda\c{s}\~{a}o para a Ci\^encia
e a Tecnologia (FCT) Scientific Employment Stimulus contract with
reference CEECINST/00032/2018, and funding from the research grants
UIDB/04434/2020 and UIDP/04434/2020.

\end{document}